\documentclass[conference]{IEEEtran}
\IEEEoverridecommandlockouts

\usepackage{cite}
\usepackage{amsmath,amssymb,amsfonts,calc}
\usepackage{algorithmic}
\usepackage{graphicx}
\usepackage{textcomp}
\usepackage{xcolor}

\usepackage{tcolorbox}
\usepackage{multirow}
\usepackage{url}
\usepackage{pgfplots}
\pgfplotsset{compat=1.18}
\usepackage{textcomp}
\usepackage{placeins}
\usepackage{fontawesome}
\usepackage{comment}
\usepackage{tabularx}
\usepackage{enumitem}
\usepackage{booktabs}

\renewcommand{\arraystretch}{1}
\usepackage{makecell}
\usepackage{csquotes}
\usepackage{multicol}
\usepackage{colortbl}
\usepackage{fancybox,framed}
\usepackage{soul}
\usepackage{tikz}
\usepackage[hidelinks,bookmarks=false]{hyperref}
\usepackage{orcidlink}

\def\BibTeX{{\rm B\kern-.05em{\sc i\kern-.025em b}\kern-.08em
    T\kern-.1667em\lower.7ex\hbox{E}\kern-.125emX}}
\begin{document}

\makeatletter
\newcommand{\linebreakand}{%
  \end{@IEEEauthorhalign}
  \hfill\mbox{}\par
  \mbox{}\hfill\begin{@IEEEauthorhalign}
}
\makeatother

\newcommand\copyrighttext{%
  \footnotesize \textcopyright~2026 IEEE. This document is a preprint. Personal use of this material is permitted.
  Permission from IEEE must be obtained for all other uses, in any current or future
  media, including reprinting/republishing this material for advertising or promotional
  purposes, creating new collective works, for resale or redistribution to servers or
  lists, or reuse of any copyrighted component of this work in other works. 
}

\newcommand\copyrightnotice{%
  \begin{tikzpicture}[remember picture,overlay]
  \node[anchor=north,yshift=-12pt] at (current page.north) {\fbox{\parbox{\dimexpr\textwidth-\fboxsep-\fboxrule\relax}{\copyrighttext}}};
  \end{tikzpicture}%
}

\title{Writing Better Software Explanations: A Guideline-Based Approach}

\author{
    \IEEEauthorblockN{
        Martin Obaidi\orcidlink{0000-0001-9217-3934},
        Jean-Carl Kremser,
        Hannah Deters\orcidlink{0000-0001-9077-7486}, 
        Jakob Droste\orcidlink{0000-0001-8746-6329}, 
        Marc Herrmann\orcidlink{0000-0002-3951-3300}, 
        Kurt Schneider\orcidlink{0000-0002-7456-8323}}
    \IEEEauthorblockA{\textit{Leibniz Universität Hannover} \\
    \textit{Software Engineering Group}\\
    Hannover, Germany \\
    \{martin.obaidi, hannah.deters, jakob.droste, marc.herrmann, kurt.schneider\}@inf.uni-hannover.de}
}

\maketitle

\copyrightnotice
\vspace{-2ex}

\begin{abstract}
As software systems increasingly rely on natural-language explanations to address user-reported explanation needs in requirements communication and support, ensuring that such explanations are consistent, relevant, and well formulated remains a major challenge. Purely automatic large language model (LLM) generation often lacks reliable grounding and controllable output quality. In this paper, we present a guideline-based formulation support tool for software explanations that combines LLM-assisted text generation with an empirically derived quality guideline. The tool structures the writing process into generation, quality checking, and iterative revision, while keeping domain control with developers. We evaluated the approach in a two-phase study consisting of an interview-based developer experiment and a controlled user survey. Six industry practitioners with software development or DevOps experience formulated explanations for real explanation needs in a human-only manual condition and in a human-with-LLM-support condition. In this small-scale evaluation, tool-supported formulation was on average 24.4\% faster, although inferential analyses indicated only a trend for efficiency. In a subsequent user study with 17 participants and 204 paired comparisons, tool-supported explanations were rated significantly higher in overall satisfaction than manual explanations ($p=0.003$, $r_{rb}=0.86$). Our findings suggest potential efficiency gains and higher perceived formulation quality through guideline-driven LLM assistance. Future work should examine long-term industrial use and integration into existing development workflows.
\end{abstract}

\begin{IEEEkeywords}
requirements engineering, explainability, guideline, survey, interviews
\end{IEEEkeywords}

\section{Introduction}
\label{sec:intro}

Software systems have become indispensable across nearly all areas of professional and personal life~\cite{Koehler2013,shklovski2014}. They support everyday activities, workplace processes, and increasingly also critical decision-making tasks~\cite{li2022}. At the same time, software systems continue to grow in complexity and scope. To use such systems effectively, users need to understand how they behave and how decisions or outcomes are produced~\cite{adadi2018peeking,anders2022mentalmodels,droste2024explanations,droste2025operationaltaxonomy}. If this understanding is missing, users may develop an \emph{explanation need}~\cite{droste2024explanations,droste2025operationaltaxonomy}, which can manifest as questions, uncertainty, frustration, or rejection of the system. In requirements engineering, such explanation needs can be treated as user feedback~\cite{anders2023userfeedback,maalej2016} that points to unmet explainability or communication requirements. Creating and refining explanations is therefore also part of requirements communication, validation, and evolution, especially when feedback from app reviews~\cite{obaidi2025appfeatures} or support requests~\cite{werner2018sentiment} is used to improve how system behavior is communicated.

Developers and support teams can address explanation needs in multiple ways, for example by adapting the system, by providing documentation, or by directly responding to users with written explanations~\cite{obaidi2025explainreqs}. This paper focuses on the latter, namely natural-language explanations formulated as direct responses to user-reported issues such as app reviews or support requests. In these settings, the wording of an explanation can influence whether it is perceived as helpful and trustworthy. A well-formulated explanation can satisfy a user's explanation need and improve acceptance, whereas a poorly phrased explanation may aggravate confusion and reduce trust~\cite{chazette2021exploring,bussone2015}.

Writing individualized explanations is time-consuming. While large language models (LLMs) can generate explanatory texts automatically, unguided generation often produces explanations that are inaccurate, irrelevant, or inconsistent with actual system behavior~\cite{obaidi2025explainreqs}. In many scenarios, developers or support staff hold the domain knowledge needed for correct content, but they need support in producing explanations that are clear, consistent, and efficient. We address this challenge by combining human content control with LLM-based writing support guided by empirically grounded quality criteria.

We operationalize these criteria in a guideline-driven formulation tool that supports three iterative steps: generating an initial explanation, checking it against guideline properties, and revising it using targeted quick fixes and rephrasing. The tool is designed as a human-in-the-loop assistant that keeps content control with the developer while the LLM supports wording and structure.

We evaluate the approach in a two-stage study. First, we conduct a task-based study with six industry practitioners with software development or DevOps experience that includes a semi-structured interview. We assess formulation efficiency and collect usability feedback while participants create explanations in a human-only manual condition and in a human-with-LLM-tool-support condition. The resulting explanations serve as stimuli for the second stage. Second, we run an independent online user study with 17 participants to compare the perceived quality of manual and tool-supported explanations in paired evaluations. Our results indicate lower average formulation time with tool support in this small-scale setting and significantly higher satisfaction ratings for tool-supported explanations.

The main contributions of this paper are:
\begin{itemize}
    \item a guideline-driven, LLM-supported formulation tool for software explanations that combines generation, quality checking, and iterative revision,
    \item a formulation guideline that supports developers and requirements engineers in translating user-reported explanation needs into structured explanation artifacts,
    \item an exploratory empirical evaluation of the tool's effect on formulation efficiency in a developer study,
    \item an empirical evaluation of the perceived quality of tool-supported versus manually written explanations in a user study,
    \item a replication package including the tool implementation, datasets, analysis scripts, and study materials to support reuse and follow-up research~\cite{obaidi2026guidelineExpDataset}.
\end{itemize}

The rest of this paper is structured as follows: Section~\ref{sec:background} reviews relevant background and related work. The study design is presented in Section~\ref{sec:research}. Section~\ref{sec:guideline-derivation} describes the derivation of the formulation guideline. Section~\ref{sec:tool-evaluation} presents the tool-based evaluation. The findings are analyzed and discussed in Section~\ref{sec:discussion}. Finally, conclusions are drawn in Section~\ref{sec:conclusion}.

\section{Background and Related Work}
\label{sec:background}

\subsection{Software Explainability}
\label{sec:software-explainability}

Software explainability has become a prominent topic in requirements engineering and software engineering research~\cite{obaidi2025elicitation,chazette2020explainability,kohl2019explainability,droste2026immersive,droste2026misunderstandings,obaidi2025automatingexplanationneedmanagement,deters2025explanationcatalog,deters2024qualitymodel,obaidi2026guidelineReq}. Explanation needs are highly individual and depend on users' goals, prior knowledge, mood, and situational context~\cite{ramos2021modeling,obaidi2025mood,obaidi2025appKonwledge,obaidi2025appfeatures}. Explainability involves multiple stakeholder groups because end users, engineers, and legal professionals often require different explanation types and presentation forms~\cite{kohl2019explainability}. This variability makes it difficult to design explainable systems that satisfy diverse expectations.

Prior work also points to trade-offs with other quality aspects. Explanations can introduce additional information that may reduce usability when not aligned with users' needs~\cite{chazette2021exploring,chazette2020explainability}. Chazette et al.~\cite{chazette2020explainability} therefore recommend user-centered design techniques and emphasize balancing costs and benefits of providing explanations. Köhl et al.~\cite{kohl2019explainability} similarly argue that explainability requirements should be treated as explicit trade-offs during development.

To support structuring and operationalization of explanation needs, several taxonomies have been proposed~\cite{unterbusch2023explanation,droste2024explanations,droste2025operationaltaxonomy}. Unterbusch et al.~\cite{unterbusch2023explanation} derived a taxonomy from app reviews. Droste et al.~\cite{droste2024explanations,droste2025operationaltaxonomy} proposed a taxonomy for everyday software systems based on an online survey. Speith~\cite{speith2022XAITaxonomies} reviewed existing explainability taxonomies and provided recommendations for their development. Chazette et al.~\cite{chazette2022can} further identify key activities for developing explainable systems and highlight that established user-centered practices can support elicitation and implementation.

\subsection{Quality Properties of Explanations and User Perception}
\label{sec:quality-properties-related-work}

Lombrozo~\cite{lombrozo2016} examined explanation preferences from a cognitive perspective and found that people prefer explanations that are simple while still providing sufficient coverage. Although not software-specific, this provides a rationale for properties related to simplicity and comprehensibility.

Work on community question answering and technical Q\&A highlights observable properties linked to perceived explanation quality. Hadfi et al.~\cite{hadfi2022} show that shallow features such as answer length and readability, and textual contents such as thematic coherence, contribute to identifying best answers. Studies in Stack Overflow contexts further emphasize tone and presentation. Calefato et al.~\cite{F.Calefato.2015} highlight the role of sentiment and presentation cues such as URLs or code snippets. Eskandari et al.~\cite{eskandari2015} report that technical terms and code snippets can be associated with higher perceived quality in technical domains. These findings support our focus on properties such as readability, clarity, relevance to the explanation need, and structured presentation. Obaidi et al.~\cite{obaidi2026usefulness} show that the perceived usefulness of developer explanations on Stack Overflow is mainly associated with structural and contextual factors such as explanation length, code inclusion, timing, and author reputation, whereas sentiment polarity has negligible influence. This supports our focus on properties that improve clarity, substance, and structured communication.

In explainability research, Deters et al.~\cite{deters2025metrics} propose criteria and metrics for assessing explainability and emphasize that assessment depends on the evaluation goal. This aligns with our use of an empirically grounded guideline that is meant to guide formulation while allowing context-dependent adaptation.

\subsection{LLMs for Explanations and Quality Control}
\label{sec:llm-authoring-related-work}

LLMs are increasingly used to generate or assist explanations, but prior work highlights both potential and risks. Lubos et al.~\cite{lubos2024} evaluate personalized LLM-generated explanations in recommender systems and report positive perceptions regarding understandability and trust. Li et al.~\cite{li2024} compare LLM explanations to expert explanations and show that LLM outputs can be preferred, often due to more concise formulations. They also indicate that instruction and strategy guidance can improve explanation performance.

At the same time, LLM-generated explanations can be incorrect. Kabir et al.~\cite{kabir2024} report that LLM-generated answers to Stack Overflow questions often contain incorrect information, yet users may still prefer them due to their well-written language. This motivates human-in-the-loop approaches that keep content control with knowledgeable staff while using LLMs for wording support.

Obaidi et al.~\cite{obaidi2025explainreqs} similarly show that automatically generated software explanations may be appreciated for clarity, tone, and style, but can still suffer from relevance and correctness issues. Their results further support hybrid approaches in which LLMs assist with drafting while humans validate factual adequacy and alignment with the explanation need.

Prompting quality directly influences generated explanations. Wiegreffe et al.~\cite{wiegreffe2022} show that prompt design affects explanation quality and that model-generated explanations can be preferred in direct comparisons under certain conditions. These findings motivate our guideline-driven prompting and hybrid quality checking, where measurable properties such as readability can be checked deterministically and other properties are assessed with targeted checks.

\subsection{Positioning of This Work}
Prior work provides taxonomies for explanation needs, quality criteria for explanations, and evidence on LLM-generated explanations. Our work complements these strands by treating the formulation of explanations as an RE-relevant communication activity: user-reported explanation needs are translated into concrete explanation artifacts, supported by an empirically derived guideline and a human-in-the-loop LLM tool.

\section{Study Design}
\label{sec:research}

\begin{figure*}[htbp]
    \centering
    \includegraphics[width=0.8\textwidth]{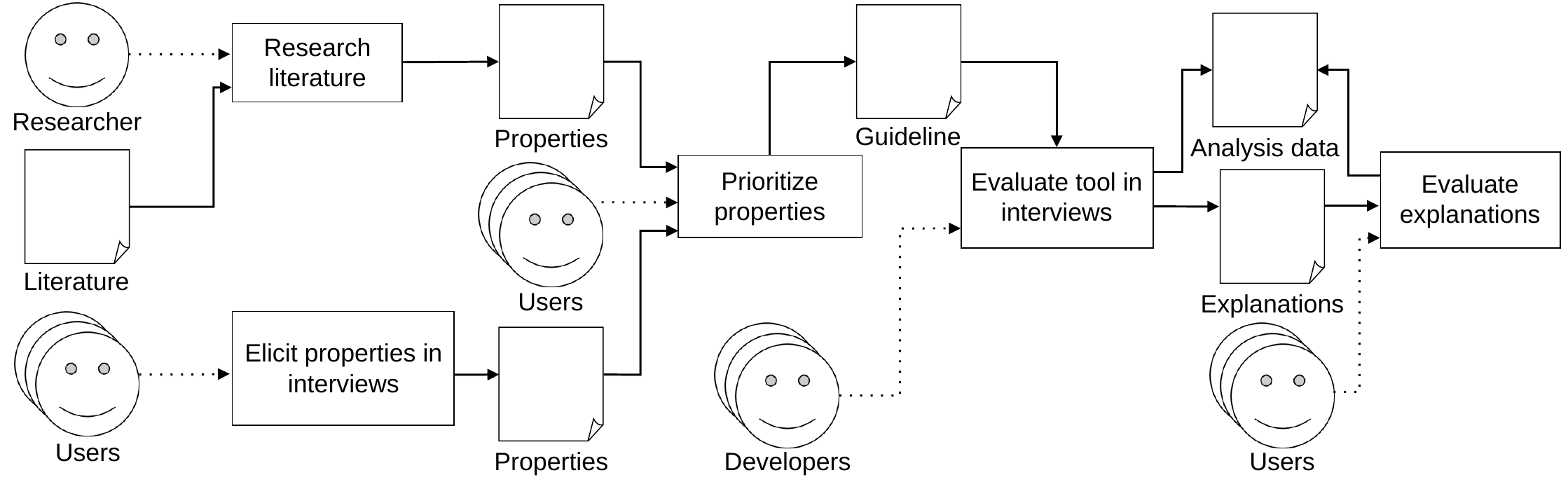}
    \caption{Overview of our sequential study design in FLOW notation~\cite{stapel2009flow}.}
    \label{fig:overview-study}
\end{figure*}

\subsection{Research Goal and Questions}
\label{sec:research-goal-questions}

The goal of this study is to evaluate a guideline-driven, LLM-supported formulation approach and tool for software explanations. We investigate which quality properties characterize good software explanations and which of these properties users perceive as most useful for formulation. We then assess the tool's impact on formulation efficiency and perceived explanation quality by comparing human-only manual writing with human writing supported by the tool from the perspective of developers and end users.

To structure our sequential study from eliciting candidate properties to prioritizing them into a guideline and evaluating its tool-based operationalization, we address the following research questions:

\begin{itemize}
    \item RQ1: Which properties characterize high-quality software explanations from research and user perspectives?\\
    This question establishes a consolidated property set that serves as the foundation for deriving an explanation formulation guideline.

    \item RQ2: Which of these properties are perceived by users as most useful for formulating software explanations?\\
    This question prioritizes the elicited properties to identify a concise set of core properties to include in the guideline.

    \item RQ3: To what extent does a guideline-based formulation tool affect the creation of software explanations compared to manual formulation in terms of efficiency and perceived quality?\\
    This question evaluates the tool-supported formulation process as an intervention and assesses how outcomes differ from human-only manual work.
\end{itemize}

Figure~\ref{fig:overview-study} outlines our sequential four-phase study design. We (1) elicit candidate properties for high-quality software explanations through a structured literature review and end-user interviews, (2) prioritize these properties in an online survey, (3) synthesize the prioritized results into a concise formulation guideline, and (4) operationalize and evaluate the guideline in a web-based formulation tool via a developer study and a subsequent online user study. Phases~1--3 are detailed in Section~\ref{sec:guideline-derivation}, and Phase~4 is reported in Section~\ref{sec:tool-evaluation}.

\subsection{Data Analysis}
\label{sec:data-analysis}

\subsubsection{Metrics and statistical procedures}
\label{sec:metrics}

\paragraph{Efficiency}
We measured formulation time per explanation in seconds. Since time measurements were right-skewed, we analyzed them using a linear mixed-effects model on log-transformed times. The model accounts for repeated measures because each developer created multiple explanations and each explanation item appears in both conditions. We included random intercepts for developer and explanation item because some developers may generally write faster or slower, and some explanation items may generally be easier or harder to formulate. As a sensitivity analysis, we additionally report a paired Wilcoxon signed-rank test on item-level manual vs.\ tool pairs.

\paragraph{Perceived quality}
Participants rated overall satisfaction with explanations on an ordinal Likert scale. For analysis, we treated the data as ordinal. For each participant and condition, we aggregated ratings by taking the median across the explanation pairs. We compared paired condition medians (Tool vs.\ Manual) using two-sided Wilcoxon signed-rank tests. We report the rank-biserial correlation ($r_{rb}$) as an effect size for paired ordinal comparisons and the Hodges--Lehmann estimator of the paired median difference ($\Delta_{HL}$) for $\mathrm{Tool} - \mathrm{Manual}$.

\paragraph{Preferences and qualitative feedback}
In addition to Likert ratings, we analyzed preference choices (manual vs.\ tool vs.\ tie) using descriptive counts. Developer feedback was summarized thematically. User comments were manually coded against the guideline property categories to connect qualitative perceptions to explanation quality properties.

\section{Guideline Derivation}
\label{sec:guideline-derivation}

This section describes how we derived the guideline for formulating high-quality software explanations in a methodologically clean way, separating methodology from results. The derivation followed a sequential four-phase design consisting of (i) elicitation of candidate properties, (ii) prioritization, (iii) guideline construction, and (iv) subsequent validation and tool-based operationalization (reported later). For the guideline derivation itself, we focus on the first three phases.

\subsection{Method}
\label{sec:guideline-method}

\subsubsection{Property Elicitation}
\label{sec:property-elicitation-method}

The goal of the first phase was to identify theoretically grounded and user-centered properties that influence the quality of software explanations. To capture both research evidence and expectations of real users, we combined a structured literature review with a semi-structured interview study.

\paragraph{Structured literature review}
The literature review was conducted in March 2025 and aimed to identify properties that influence the perceived quality of explanations. Because only a few studies explicitly address \emph{software explanations}, we deliberately broadened the search scope to related domains, including Q\&A communities, comprehensibility and readability research, explanation quality in general, and explainability.

We included papers that identify at least one concrete property affecting explanation quality and describe it in sufficient detail to be operationalized in an LLM-based tool. We further required that the property is transferable to software explanations in principle. We excluded papers that were not accessible via university resources or not available in German or English.

The search was performed using Google Scholar and the AI-supported search engine Consensus\footnote{\url{https://consensus.app/}} with multiple keyword combinations, for example ``properties of good software explanations'' combined with \textit{properties}, \textit{features}, \textit{qualities}, \textit{answers}, and \textit{responses}. In total, we screened 100 publications, read 30 in full, and selected 11 as relevant. From these, we extracted the initial set of explanation properties.

\paragraph{User interviews}
In a second step, we conducted online interviews in April 2025 as an exploratory complement to the literature-based properties and to collect user-centered examples.

\paragraph{Participants}
Three participants aged 21--25 took part in the interviews. The sessions were conducted online and lasted between 63 and 93 minutes. None of the participants worked in IT. However, all reported recurring needs for software-related explanations.

\paragraph{Procedure}
The interviews followed a semi-structured format. We began with an introduction that explained the study goal, procedure, and the concepts of \emph{explanation need} and \emph{explanation}. Participants then provided basic demographic information and described their usage context. To create a shared understanding, we presented selected app review posts~\cite{obaidi2026multigolddataset} and five example explanations. Based on these materials, participants evaluated the example explanations and articulated what they consider important for good software explanations. They then identified desired properties, grouped mentioned aspects into broader categories (e.g., tone, simplicity, readability), and finally named properties that should be avoided in good explanations. We concluded each session with an open reflection.

\paragraph{Analysis}
Interview transcripts were manually corrected, coded, and categorized by one coder. In a first coding iteration, we identified (1) higher-level properties of good software explanations and (2) \emph{concretizing mentions} used by participants to describe, refine, or exemplify these properties (e.g., ``paragraphs after the greeting'' or ``a separate paragraph for links'' as concretizations of the property \emph{paragraphs}). At this stage, these concretizations served to refine the semantic meaning of properties and were not yet interpreted as independent property categories.

Overall, the first coding pass yielded 30 properties and 47 concretizing mentions. Semantically similar mentions were merged and assigned to overarching properties already during this phase.

\subsubsection{Property Prioritization Survey}
\label{sec:property-prioritization-method}

\paragraph{Preparation and consolidation}
For the prioritization study, we consolidated the results from the literature review and the user interviews into an operationalizable property list. The goal of this phase was to assess the relevance of the identified properties from a user perspective and to provide an empirical basis for guideline derivation.

To this end, semantically similar properties were merged, and the concretizing mentions collected in the interviews were condensed into \emph{property manifestations} that could be rated unambiguously in a survey. At the same time, properties that were not applicable in the target scenario were excluded (e.g., \emph{images}, since this work focuses on purely textual explanations). In addition, some properties that had initially been coded separately were modeled as manifestations of a higher-level property (e.g., \emph{politeness}, \emph{factuality}, or \emph{friendliness} as manifestations of \emph{writing style}).

This consolidation resulted in 21 higher-level properties, seven of which were split into multiple manifestations for more precise assessment (29 in total). Altogether, the survey therefore comprised 43 rateable property manifestations.

\paragraph{Participants and recruitment}
The prioritization was conducted as an online survey implemented in LimeSurvey. The survey was available from May 18 to May 28, 2025, via a LimeSurvey link and used convenience sampling through personal networks and student channels. We deliberately did not restrict participation to a narrow target group, as in principle any user may experience software-related explanation needs. In addition to the property ratings, we collected demographic information and self-reported software experience, as well as prior experience with software-related explanation needs.

\paragraph{Survey instrument and procedure}
The prioritization study was conducted as an online survey. It started with a short introduction explaining the survey goals and defining the concepts of \emph{software explanation} and \emph{explanation need}, illustrated with examples. Participants then completed a demographics section covering gender, age, software experience, and prior explanation needs in software contexts. In the main part, participants rated all 43 candidate property manifestations of good software explanations consolidated from the literature review and interviews. To reduce cognitive load, we grouped related properties into broader categories where appropriate, such as tone, clarity, and readability. All candidate properties were rated on a Likert scale for perceived usefulness.

\paragraph{Prioritization logic}
For the analysis, Likert responses were recoded to a scale from \(-3\) to \(+3\), with \(0\) representing the neutral midpoint. Since Likert ratings are ordinal, we used the median as the primary decision criterion for prioritization. Properties were classified as follows:

\begin{itemize}
    \item Median \(\ge 2\): property considered useful and included in the guideline.
    \item Median \(\le -2\): property considered not useful and included as a negative recommendation (i.e., a property to avoid).
    \item Median between \(-2\) and \(2\): property considered neutral and not included in the guideline core, but potentially relevant depending on context.
\end{itemize}

This design ensured that the resulting guideline reflects properties that are both theoretically grounded and empirically perceived as relevant for formulating good explanations.

\subsubsection{Guideline Construction}
\label{sec:guideline-construction-method}

\paragraph{Selection of the final property set}
Based on the prioritization results, we included those properties in the guideline that were rated predominantly positively in the survey (Median \(\ge 2\)). If a property had been operationalized through multiple manifestations in the survey, only the relevant manifestations were translated into concrete recommendations in the guideline.

Properties with a median of \(\le -2\) were incorporated as \emph{properties to avoid}, including, for example, humorous or romantic writing styles, the use of emojis, and irrelevant advertising. Properties with neutral median values were not included in the guideline core because participants rated them inconsistently and thus they did not provide a sufficiently clear empirical basis for a core recommendation.

Overall, this yielded a set of 22 positively rated manifestations translated into guideline recommendations, forming the backbone of the guideline.

\paragraph{Guideline structure}
The selected properties were then consolidated into a guideline describing central properties of high-quality software explanations. The resulting artifact is structured as a table with the columns \emph{Property}, \emph{Definition}, and \emph{Recommendation}. This structure was chosen to support both human use (as a concise writing aid) and later operationalization in LLM prompts and quality checks.

\paragraph{Scope and transferability considerations}
The guideline was derived primarily in the context of user-directed responses to explanation needs expressed in app reviews. Such responses are often created outside the actual software system (e.g., by support teams) and therefore do not constitute system-internal explanations in a narrow sense. We nevertheless treat them as a relevant explanation format because they directly address explanation needs in practice and are subject to the same quality demands of understandability and need fit.

At the same time, explanation needs can also be addressed through system-internal explanations (e.g., tooltips, contextual help, tutorials, or question-mark icons in the user interface) and through external documentation (e.g., FAQs or manuals). We therefore considered the broader transferability of the identified properties across explanation formats. In particular, some properties are more relevant for dialogic responses than for impersonal UI text (e.g., greeting/closing formulas or explicit requests for feedback), while others remain broadly applicable across formats (e.g., clarity, relevance, readability, and structuring).

\subsection{Results}
\label{sec:guideline-results}

\subsubsection{Elicited Candidate Properties}
\label{sec:elicited-properties-results}

This section reports the results of the property elicitation phase, combining findings from the structured literature review and the user interviews.

\paragraph{Literature review results}
The analysis of the 11 selected publications yielded four overarching categories of explanation properties relevant across different domains:

\textit{Sentiment.}
Studies such as Kucuktunc et al.~\cite{kucuktunc2012}, Ferguson et al.~\cite{ferguson2024}, and Calefato et al.~\cite{F.Calefato.2015} indicate that the emotional tone of an explanation can influence how it is perceived, including its persuasiveness and perceived humanness.

\textit{Writing style.}
Work by Okoso et al.~\cite{okoso2025}, Lee et al.~\cite{lee2019}, and Blooma et al.~\cite{blooma2012} suggests that politeness, attractive language, and clear wording can increase user satisfaction and trust.

\textit{Simplicity.}
Studies by Sehl et al.~\cite{sehl2024}, Zemla et al.~\cite{zemla2017}, and Lombrozo~\cite{lombrozo2016} emphasize the importance of simplicity. At the same time, they indicate that the preferred level of simplicity depends on context, i.e., users may prefer simpler or more complex explanations depending on the situation.

\textit{Readability.}
Research such as Withall et al.~\cite{withall2021readability}, Blooma et al.~\cite{blooma2012}, Hadfi et al.~\cite{hadfi2022}, and again Zemla et al.~\cite{zemla2017} highlights the influence of length, comprehensibility, readability level, and linguistic clarity on trust and understandability.

These categories served as the theoretical foundation for the subsequent interview study.

\paragraph{Interview results}
The interviews provided user-centered examples that were consistent with many literature-based properties. In the first coding iteration, we identified 30 properties and 47 concretizing mentions. Frequently mentioned desirable properties included clarity, brevity/conciseness, relevance, politeness, simplicity, logical structure, and respectful address. Participants also explicitly identified properties that should be avoided in good explanations, most notably emojis, advertising, informal language, and unnecessary filler phrases.

Overall, the interview findings enriched the literature-derived categories with concrete user-oriented expectations and examples that informed the subsequent consolidation and prioritization steps.

\paragraph{Synthesis of literature and interviews}
Taken together, the literature review and user interviews yielded an initial candidate set of explanation quality properties that was both theoretically grounded and informed by actual user expectations. The literature contributed broad conceptual categories (e.g., sentiment, writing style, simplicity, readability), while the interviews refined these into more concrete, application-oriented properties and highlighted negative examples to avoid.

This combined set formed the basis for the subsequent consolidation into an operationalizable property list for the prioritization survey (reported in Section~\ref{sec:prioritized-properties-results}).

\subsubsection{Prioritized Properties}
\label{sec:prioritized-properties-results}

\paragraph{Participants}
The online survey included 17 participants, of whom 65\% identified as male and 35\% as female. Most participants (88\%) were between 20 and 39 years old. About 35\% reported that they had previously expressed at least one explanation need in a software context. Regarding software knowledge, 59\% rated themselves as good or very good, 29\% as intermediate, and 12\% as rather low. No participant reported having no software experience.

\paragraph{Property ratings}
This quantitative survey served to systematically evaluate all previously identified properties. Figure~\ref{fig:survey-diverging-bars} shows the distribution of Likert responses across all rated property manifestations.

\begin{figure*}[htbp]
  \centering
  \includegraphics[width=0.85\textwidth]{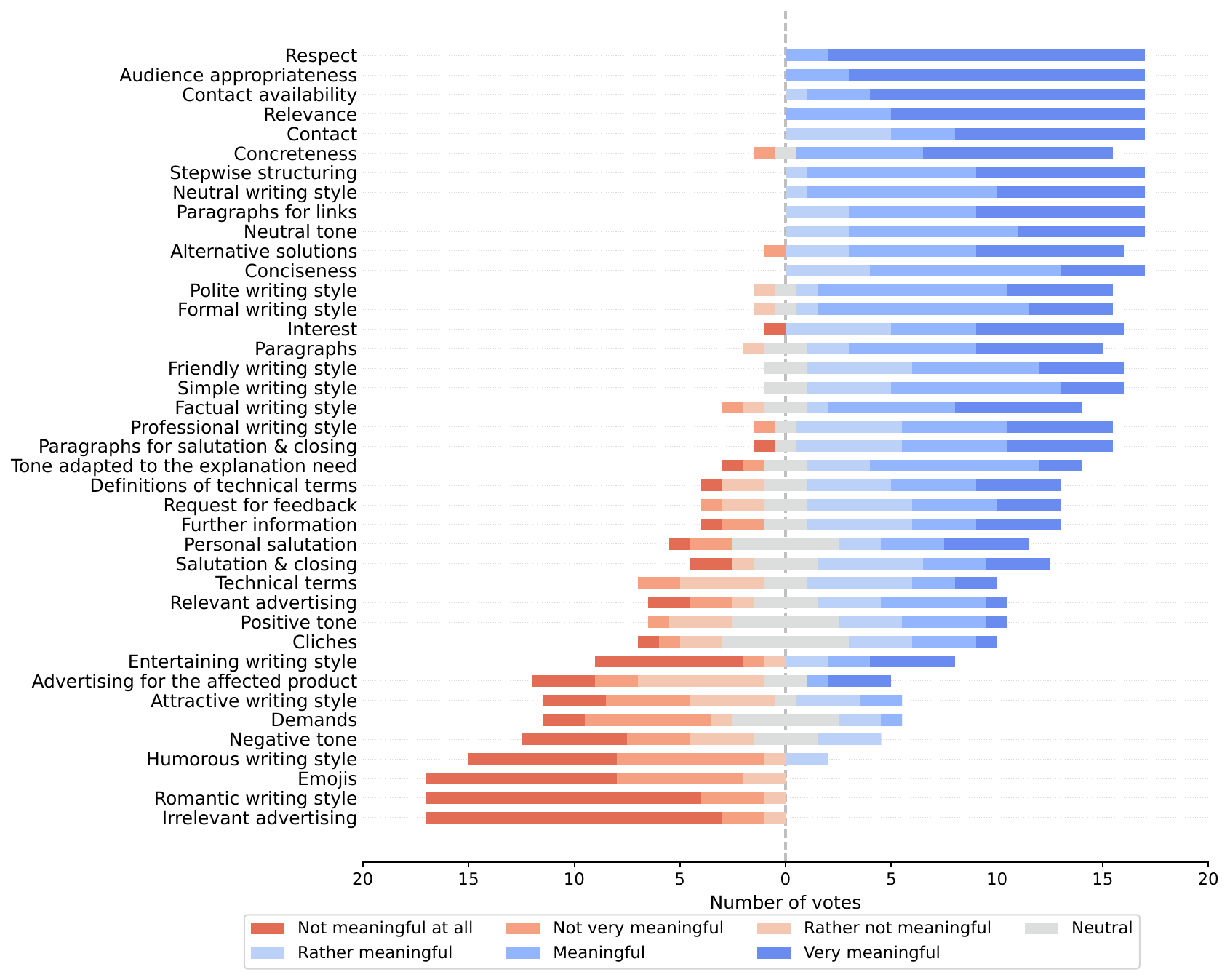}
  \caption{Likert ratings of all property manifestations (n=17), sorted by median}
  \label{fig:survey-diverging-bars}
\end{figure*}

\begin{table*}[htbp!]
\centering
\scriptsize
\renewcommand{\arraystretch}{1.3}
\caption{Guideline for formulating high-quality software explanations: definitions and concrete formulation instructions to be applied during explanation writing}
\label{tab:kremser-guideline}
\begin{tabularx}{\textwidth}{@{}p{2.8cm} p{6.0cm} X@{}}
\toprule
\textbf{Property} & \textbf{Definition} & \textbf{Recommendation} \\
\midrule
\textbf{Paragraphs} &
Visible text breaks that structure a text and mark thematic units. &
Paragraphs should be used frequently. Greeting, closing, and links should be separated into their own paragraphs. \\
\textbf{Audience appropriateness} &
Adapting the explanation to the user's prior knowledge and language abilities. &
The explanation should be formulated in an audience-appropriate way. \\
\textbf{Emojis} &
Symbols such as smileys or pictograms that visualize emotions or content. &
Emojis should generally not be used. \\
\textbf{Contact} &
Providing the responsible person or team and a way to give feedback. &
A contact should always be provided, preferably a team or person name, supplemented by a feedback option. \\
\textbf{Concreteness} &
Specific, unambiguous, and tangible wording instead of vague or generic statements. &
All information should be clear and concrete. \\
\textbf{Brevity \& conciseness} &
Condensed presentation focused on the essential content. &
The explanation should be short and concise. \\
\textbf{Readability} &
The ease with which a text can be read due to its linguistic structure. &
The Flesch index of the explanation text should be in the range 40--49, corresponding roughly to newspaper-level language. \\
\textbf{Relevance} &
The degree to which the explanation content fits the expressed explanation need. &
The explanation should be relevant to the explanation need. \\
\textbf{Writing style} &
The manner of linguistic expression, e.g., formal, friendly, or factual. &
A neutral writing style is recommended. Formal, simple, friendly, professional, factual, or polite variants may be combined depending on context. \\
\textbf{Structuring} &
Logical and comprehensible organization of information into suitable steps/sections. &
The explanation should be structured in a stepwise and logical manner. \\
\textbf{Tone} &
The emotional tone of the text, from neutral to adapted to the explanation need. &
The explanation should be formulated neutrally. Alternatively, it may be adapted to the tone of the explanation need. \\
\textbf{Advertising} &
Statements or references intended to promote products or services. &
Advertising should be avoided entirely, especially when it does not contribute to solving the explanation need. \\
\bottomrule
\end{tabularx}
\end{table*}

Overall, 22 property manifestations were classified as useful (Median \(\ge 2\)). Four manifestations were rejected and later incorporated into the guideline as negative recommendations: \textit{humorous writing style}, \textit{romantic writing style}, \textit{emojis}, and \textit{irrelevant advertising}.

A further central finding concerned appropriate readability. The survey results showed that users preferred explanations with a Flesch readability index in the range of 40--49, corresponding roughly to newspaper-level language. This range provides an empirically grounded target value for later automated generation and revision procedures in our study context.

In addition, participants preferred regular use of paragraphs, which support structuring and comprehensibility of explanations. Providing a concrete contact person or team was rated as helpful by a majority of participants, whereas naming only a company was considered less useful.

Overall, the results indicate that both structural properties (e.g., paragraphing, readability, clarity) and social aspects (e.g., contact information) contribute to the perceived quality of software explanations. At the same time, the findings show that certain stylistic properties are perceived as inappropriate in the software explanation context.

\subsubsection{Final Guideline}
\label{sec:final-guideline-results}

Based on the prioritization results, we derived a guideline for formulating high-quality software explanations. The guideline combines positively prioritized properties as explicit recommendations and negatively prioritized properties as explicit ``avoid'' instructions. In total, 22 positively rated property manifestations were translated into recommendations.

The guideline indicates that explanations should be particularly relevant, concise, clear, neutral in tone, and well structured. As a readability target, a Flesch index of 40--49 was identified, corresponding roughly to newspaper-level language and supporting understandable yet precise wording. Writing style should be neutral or professional depending on context: simple, friendly, and polite formulations are also appropriate. Frequent paragraphing supports orientation, and explicit contact information, preferably naming a team or person, was rated as helpful. In contrast, emojis, humorous or romantic styles, and advertising should be avoided.

Table~\ref{tab:kremser-guideline} shows the resulting guideline with properties, definitions, and concrete formulation recommendations.

Although the guideline was primarily derived for user-directed responses to app-review-based explanation needs, many of its properties are also applicable to other explanation formats. Nearly all properties remain potentially relevant for user-directed responses, except for practical constraints such as \emph{images} in purely textual channels. For system-internal explanations, however, some properties are mainly relevant in dialogic settings and less applicable to impersonal UI text (e.g., greeting/closing formulas or explicit requests for feedback). Overall, the guideline is intended as a transferable starting point, with context-specific adaptation depending on explanation format.

\section{Tool Development and Evaluation}
\label{sec:tool-evaluation}

\subsection{Tool Overview}
\label{sec:tool-overview}

We operationalized the derived guideline in a web-based formulation support tool for software explanations. The tool is designed to support developers during writing rather than to fully automate explanation creation. This design follows our earlier observation that purely automatic LLM generation often struggles with relevance and factual adequacy in software explanation contexts.

We briefly describe the workflow, prompting strategy, and core modules that are relevant for the evaluation. The tool, prompt templates, datasets, analysis scripts, and additional materials are available in our replication package~\cite{obaidi2026guidelineExpDataset}.

Figure~\ref{fig:tool-screenshot} shows the main interface of our web-based formulation support tool.

\begin{figure*}[htbp]
    \centering
    \includegraphics[width=0.80\textwidth]{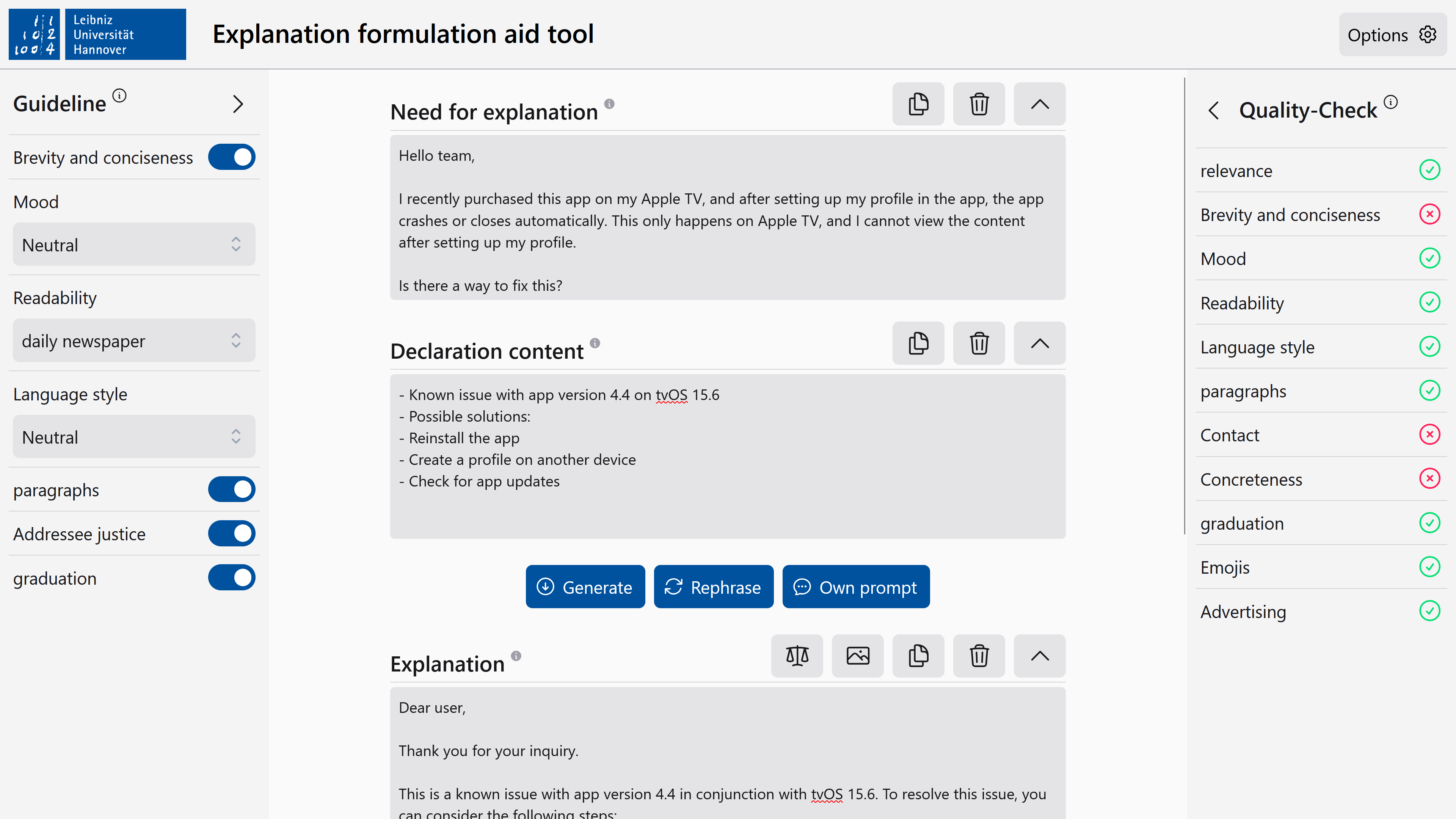}
    \caption{Screenshot of the tool showing guideline configuration, explanation editing, and property-level quality checking.}
    \label{fig:tool-screenshot}
\end{figure*}

\paragraph{Conceptual workflow}
The tool supports a three-step workflow that uses the guideline as a consistent control mechanism across all stages.

\begin{enumerate}
    \item \textbf{Initial generation.} Based on an \emph{explanation need} and optional \emph{explanation content}, the LLM generates an initial explanation draft. The optional content serves as domain grounding and reduces the risk of incorrect or hallucinated statements. Guideline properties are injected in compact form through the system instruction and steer style, structure, and relevance.

    \item \textbf{Quality checking.} The generated explanation is assessed against the guideline properties. Content- and language-related properties are checked via LLM-based quality-check prompts. Measurable properties such as readability and paragraph structure are checked with deterministic functions. The tool provides property-level feedback to make strengths and weaknesses transparent.

    \item \textbf{Iterative revision.} The explanation can be revised based on feedback. Revision support includes quick fixes (targeted LLM instructions), manual edits, rephrasing, and custom prompts for targeted changes.
\end{enumerate}

This iterative process gradually aligns the explanation with the guideline. Since LLM outputs are non-deterministic, quality checks and quick fixes serve as pragmatic support rather than guarantees.

\paragraph{Prompting strategy}
The prompting design is modular and separates prompts for generation, quality checks, and revision. A compact system instruction encodes the guideline properties and provides the common behavioral frame for all interactions. Generation prompts create an explanation from the explanation need and optional grounding content. Quality-check prompts evaluate the current explanation property by property and return short feedback messages. Quick-fix prompts apply targeted revisions for detected issues.

\paragraph{Architecture and modules}
The tool was implemented as a web application with a modular structure. LLM integration uses the OpenAI API with \textit{GPT-4o-mini}. API calls are encapsulated in a separate service layer to simplify later model replacement or extension.

The desktop-oriented user interface follows the writing workflow and consists of three main areas: guideline configuration, formulation, and quality checking.

The \emph{generation module} combines the explanation need, optional explanation content, and current guideline settings into a structured prompt and displays the generated explanation as editable text with optional Markdown rendering. The \emph{guideline module} allows users to activate or deactivate individual guideline properties. Deactivated properties are removed from the instruction context.

The \emph{quality-check module} implements a hybrid evaluation strategy. Most properties are assessed with LLM-based checks. Readability is evaluated deterministically via a Flesch index. Paragraph structure is assessed deterministically via a minimum number of paragraphs per 100 words because these criteria were not reliably handled by the LLM during development. The module highlights fulfilled and violated properties and supports revision through quick fixes.

\subsection{Evaluation Method}
\label{sec:tool-evaluation-method}

We evaluated the guideline-driven formulation tool in two complementary phases. First, we conducted an interview-based developer study to assess formulation efficiency and collect usability feedback. Second, we conducted an independent online user study to assess the perceived quality of human-only manual and human-with-LLM-tool-supported explanations. The participants in this online user study were recruited independently from the participants in the property prioritization survey. The tool was fully implemented and configured with the guideline derived in Section~\ref{sec:guideline-derivation}. All participants were at least 18 years old. The online survey was anonymous.

\subsubsection{Developer Study}
\label{sec:developer-study-method}

\paragraph{Participants and recruitment}
We recruited six industry practitioners with software development or DevOps experience from TIXEL GmbH. Participants were between 29 and 44 years old. All were male. Five participants worked primarily in software development and one participant worked in DevOps and IT support. Professional experience ranged from 3 to 15 years.

\paragraph{Procedure and measures}
Each participant worked on six real \emph{explanation needs} from a gold-standard dataset~\cite{obaidi2026multigolddataset}. For each participant, three explanations were created in a human-only manual condition and three in a human-with-LLM-tool-support condition. This yielded paired outputs across the study that were later used as stimuli in the online user study.

To improve comparability, each explanation need was accompanied by bullet-point \emph{explanation content} that served as domain grounding. This reduced content-related variation and shifted the evaluation focus to formulation and structure rather than problem solving.

Sessions were conducted via Microsoft Teams and lasted between 55 and 90 minutes. We recorded formulation time per explanation in seconds as the primary efficiency metric. In addition, we collected qualitative feedback on usability, strengths, weaknesses, and improvement suggestions in a semi-structured closing part of the interview.

\subsubsection{Online User Study}
\label{sec:online-quality-method}

\paragraph{Participants and recruitment}
The study included 17 participants. Participation required age $\ge 18$. The sample consisted of 59\% male and 41\% female participants. Most participants were between 20 and 29 years old (70\%), followed by 30--39 years (18\%) and 40+ years (12\%). Regarding education, 41\% held a Bachelor's degree, 29\% had a university entrance qualification, and 12\% held a Master's degree or equivalent. Two participants reported vocational training and one participant reported no school degree. Self-rated software knowledge was good or very good for 76\% of participants, while 18\% reported low or no familiarity.

\paragraph{Instrument and procedure}
The stimuli were paired explanations produced in the developer study, consisting of one manual and one tool-supported explanation for the same explanation need. The developer study produced 18 matched explanation pairs in total. To keep the online user study manageable and reduce participant fatigue, we randomly selected 12 of these pairs as stimuli. In total, participants evaluated 12 explanation pairs, resulting in 204 paired comparisons (\(17 \times 12\)) and 408 individual ratings.

Participants rated their overall satisfaction with the formulation of each explanation on an ordinal Likert scale. For analysis, responses were recoded to \([-3, +3]\), with \(0\) as the neutral midpoint. Participants could also indicate which version they preferred (manual, tool-supported, or equal). Two closing questions captured confidence in the ratings and the perceived importance of well-formulated software explanations.

\subsection{Results}
\label{sec:tool-evaluation-results}

\subsubsection{Efficiency Results}
\label{sec:efficiency-results}

In the developer study, participants created 36 explanations in total, comprising 18 manual and 18 tool-supported explanations. Figure~\ref{fig:duration-overall} shows the recorded formulation times per explanation pair.

\begin{figure}[htbp]
    \centering
    \includegraphics[width=0.9\columnwidth]{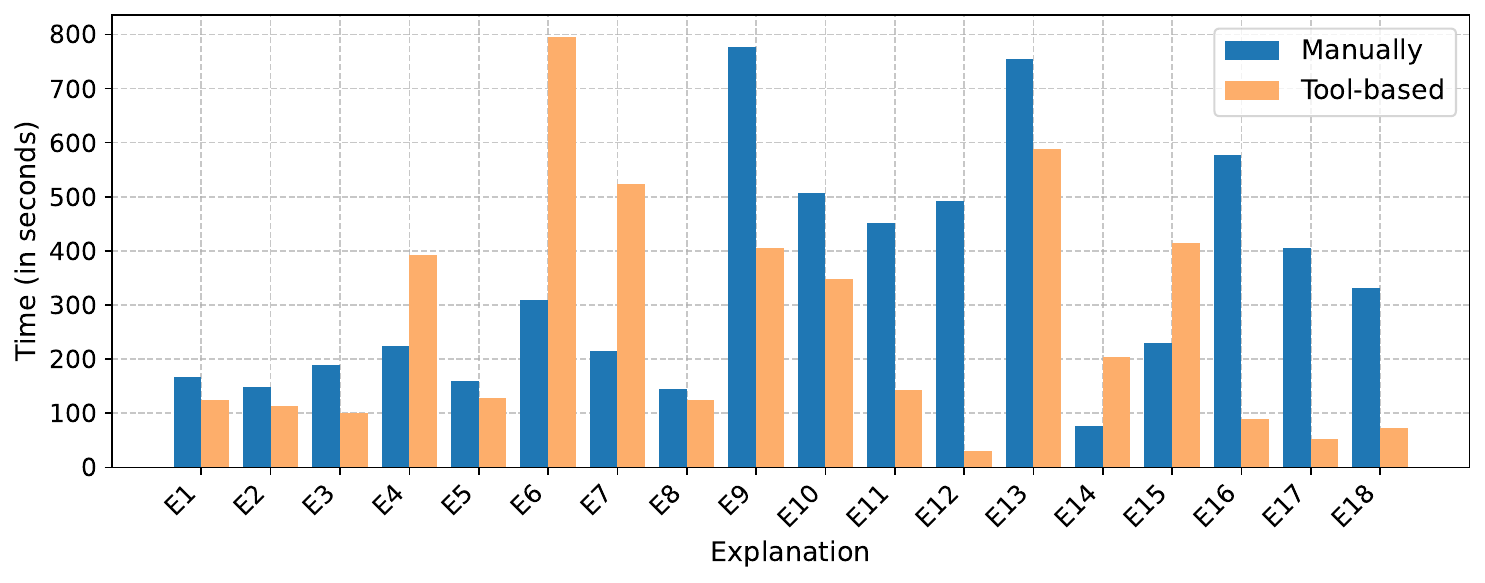}
    \caption{Formulation time for manual and tool-supported explanations}
    \label{fig:duration-overall}
\end{figure}

On average, manually formulated explanations required \(342.1\) seconds, while tool-supported explanations required \(258.7\) seconds. Descriptively, this corresponds to an average reduction of \(83.4\) seconds (\(24.4\%\)) with tool support. Pairwise differences varied substantially across explanation items. Some pairs showed only small differences, while others showed much faster tool-supported formulation. A few cases were slower with the tool, which participants attributed to initial familiarization with the interface and quality checks.

The item-level paired Wilcoxon signed-rank test indicated a trend toward shorter formulation times with tool support (one-sided: \(W=55\), \(p=0.098\); two-sided: \(p=0.196\)). A linear mixed-effects model on \(\log(\text{time})\) with random intercepts for developer and explanation item showed a similar trend, with an estimated time reduction of approximately \(36.6\%\) for tool-supported formulation (\(p=0.065\)).

Qualitative developer feedback was predominantly positive. Five of the six developers considered the tool useful, especially for recurring formulation tasks. Frequently mentioned strengths included guideline integration, quick-fix functions, and rapid rephrasing support. Reported limitations mainly concerned usability features such as version history, templates, autosave, and interaction speed. Five developers stated that they could imagine using the tool in professional practice.

\subsubsection{Perceived Quality Results}
\label{sec:quality-results}

The online user study comprised 12 explanation pairs, resulting in 204 paired comparisons (\(17 \times 12\)) and 408 individual ratings. Across all comparisons, tool-supported explanations were preferred in 117 cases, manual explanations in 54 cases, and both versions were rated equally in 33 cases.

Figure~\ref{fig:kremser-survey-2-manual-tool-Median} shows the median recoded satisfaction ratings for manual and tool-supported explanations per explanation pair.

\begin{figure}[htbp]
    \centering
    \includegraphics[width=0.95\columnwidth]{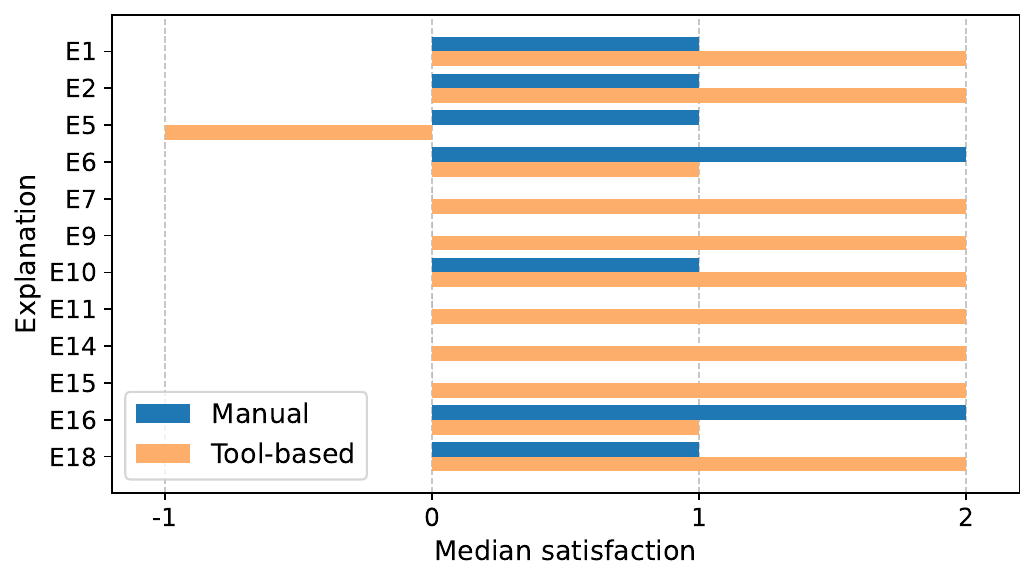}
    \caption{Median satisfaction ratings for manual vs.\ tool-supported explanations per explanation pair}
    \label{fig:kremser-survey-2-manual-tool-Median}
\end{figure}

Overall, the aggregated median satisfaction was \(1\) for manual explanations and \(2\) for tool-supported explanations. Dispersion was lower for tool-supported explanations (\(\mathrm{IQR}=1\)) than for manual explanations (\(\mathrm{IQR}=3\)), indicating more consistent ratings in favor of the tool-supported variants. At the participant level, 14 of 17 participants more often preferred tool-supported explanations across all pairs, one participant more often preferred manual explanations, and two participants rated both conditions roughly equally often. At the item level, tool-supported explanations were preferred in 9 of 12 explanation pairs and manual explanations in 3 of 12 pairs.

The two-sided Wilcoxon signed-rank test on participant-level median ratings indicated significantly higher perceived satisfaction for tool-supported explanations (\(W=8.5\), \(p=0.003\)). The effect size was large (\(r_{rb}=0.86\)). The Hodges--Lehmann estimator indicated a robust location shift of \(\widehat{\Delta}_{HL} \approx 1.5\) Likert points in favor of tool-supported explanations.

\subsubsection{Participant Confidence and Perceived Importance}
\label{sec:closing-questions-results}

Participants reported high confidence in their judgments. Overall, 94\% selected ``absolutely sure'', ``sure'', or ``rather sure'' when asked about confidence in their ratings. In addition, 88\% rated well-formulated software explanations as important or very important. These results suggest that participants felt able to judge the explanations and considered formulation quality relevant from a user perspective.

\section{Discussion}
\label{sec:discussion}

In the following, we answer the research questions, present threats to validity, and interpret the results.

\subsection{Answers to the Research Questions}
\label{sec:beantworten-der-forschungsfragen}

\textbf{RQ1: Which properties characterize high-quality software explanations from research and user perspectives?} \\
Across the literature review and exploratory user interviews, high-quality software explanations were characterized by a combination of structural, linguistic, and social properties. The literature highlighted broad categories such as sentiment, writing style, simplicity, and readability. The interviews provided user-centered examples consistent with these categories and refined them into concrete properties, especially clarity, brevity/conciseness, relevance, politeness, simplicity, logical structuring, and respectful address, while also identifying undesirable properties such as emojis, advertising, informal language, and unnecessary filler phrases.

\textbf{RQ2: Which of these properties are perceived by users as most useful for formulating software explanations?} \\
The prioritization survey indicated that participants perceived a substantial subset of the elicited properties as useful for formulating software explanations, with 22 property manifestations rated positively (Median \(\geq 2\)). The resulting guideline emphasizes relevance, clarity, brevity/conciseness, logical stepwise structuring, readable language, neutral/professional writing style, paragraphing, and explicit contact information. Participants also preferred a Flesch readability range of 40--49 as a concrete target. In contrast, humorous or romantic writing style, emojis, and irrelevant advertising were rated as not useful and were included as explicit properties to avoid.

\textbf{RQ3: To what extent does a guideline-based formulation tool affect the creation of software explanations compared to manual formulation?} \\
The evaluation provides promising evidence for higher perceived formulation quality and potential efficiency gains. In the developer study, tool-supported formulation was descriptively 24.4\% faster on average (83.4 seconds), while inferential analyses indicated only a trend toward faster formulation. In the online user study, tool-supported explanations were rated significantly higher than manual explanations in overall satisfaction (\(p=0.003\), \(r_{rb}=0.86\)), and were preferred more often at both participant and item level. Overall, the results suggest that guideline-based LLM support can improve perceived formulation quality, while conclusions about stable efficiency gains remain preliminary.

\subsection{Interpretation}
\label{sec:interpretation-results}

\paragraph{Guideline-driven assistance as a pragmatic use of LLMs}
Our results suggest that LLMs add most value when used as guided drafting support under human content control. In our setting, developers received the factual explanation content and used the tool mainly to improve wording and structure. Even with this constrained scope, the tool produced explanations that users preferred significantly more often and rated higher in satisfaction, while developers spent less time on average. This indicates that guideline-driven revision and checking can reduce writing effort and improve perceived formulation quality without relying on fully automatic generation.

\paragraph{What matters most for users in this context}
The property prioritization and the evaluation point to a small set of levers that appear especially impactful in user-facing explanations. Users consistently favored explanations that are relevant to the expressed need, structured in steps, concise, and easy to read. The survey also provides a concrete readability target (Flesch 40--49) and supports simple structural rules such as regular paragraphing. At the same time, some stylistic choices that may work in other settings were clearly rejected here, including emojis and humorous or romantic tone. For practice, this implies that improving explanations does not require a large style guide, but a focused checklist that emphasizes relevance, structure, readability, and an appropriate professional tone.

\paragraph{Implications for RE practice and research}
For requirements engineering and explainability engineering, the results imply that eliciting explanation needs is not sufficient if downstream communication remains unstructured. Teams that manage explainability requirements can use a formulation guideline as a lightweight complement when translating user feedback into explanation artifacts, especially for support responses, help texts, and documentation. For RE practitioners, the key takeaway is to separate explanation \emph{content} from explanation \emph{formulation}. Our study controlled content and still observed meaningful differences, which suggests that formulation quality is an actionable layer that can be operationalized and evaluated. This supports integrating formulation criteria into review and validation routines when explainability-related texts are part of the delivered system experience. For research, the prioritization results indicate that explanation quality is context-sensitive. Since some properties are universally preferred while others depend on the setting, this motivates future work on adaptive guideline profiles and stronger links between measurable automated checks and perceived quality outcomes.

\subsection{Threats to Validity}
\label{sec:validity}

We discuss threats to validity following Wohlin et al.~\cite{wohlin2012experimentation}.

\subsubsection{Construct Validity}
Our main constructs, namely explanation quality, property usefulness, efficiency, and perceived quality, were operationalized through Likert ratings, preference choices, and completion times. These measures capture relevant facets but do not fully cover the underlying concepts. Ratings may also vary with individual expectations about what constitutes a ``good'' explanation. In addition, results depend on the specific LLM (\textit{GPT-4o-mini}) and the prompt and check design used to operationalize the guideline. The rating-based measures may also be affected by cognitive biases. The prioritization survey captures declared usefulness, which may differ from users' actual priorities in real use situations. In the paired evaluation, fluent and professionally phrased texts may have benefited from fluency bias. Automation-related expectations may also have influenced judgments, although participants were not asked to identify whether a text was tool-supported.

\subsubsection{Internal Validity}
Differences in developers' writing habits and familiarity with the prototype likely influenced completion times. The controlled setup used bullet-point explanation content to reduce content variation, which improves comparability but does not reflect situations where authors must both solve the issue and write the explanation. In the online study, repeated pairwise judgments may have introduced fatigue or contrast effects. The user interviews were exploratory and based on a small sample, so they should be interpreted as a complement to the literature review rather than as evidence of saturation.

\subsubsection{Conclusion Validity}
Sample sizes were small (\(n=6\) developers and \(n=17\) users), which limits statistical power, especially for efficiency where results indicate trends. The 24.4\% time reduction is therefore a descriptive observation and should not be interpreted as a stable effect estimate. Likert data is ordinal, which we addressed with non-parametric tests and robust estimators, but subjective ratings can still mask individual preference structures.

\subsubsection{External Validity}
The studies used explanation needs derived from app reviews and evaluated texts outside real production settings. Results may differ for other domains, explanation formats, languages, and organizational workflows. The survey samples were recruited through convenience sampling and may overrepresent educated or technology-affine users. The developer sample came from a single company and was all male, which further limits generalizability. The literature review was structured but not a full systematic literature review, so relevant publications may have been missed. Outcomes may also change for other LLMs, interfaces, or guideline operationalizations.

\subsection{Future Work}
\label{sec:future-work}

Future work should validate the guideline-driven approach in broader settings. Our studies used explanation needs derived from app reviews, so replication across other channels such as support tickets, forums, and in-app feedback, as well as across additional domains and languages, is needed.

Methodologically, future evaluations should go beyond overall satisfaction and assess more directly whether tool-supported explanations satisfy specific guideline properties and how these properties relate to user judgments. Finally, the tool could be strengthened with more robust quality assurance for factual correctness and need fit, for example by integrating organizational knowledge sources and lightweight consistency checks. Another promising direction is context-sensitive guidance, where the tool adapts guideline emphasis and prompting to different explanation formats such as support responses, UI help texts, and documentation.

\section{Conclusion}
\label{sec:conclusion}

This paper presented a guideline-driven approach to improve the formulation of software explanations and evaluated its operationalization in an LLM-supported formulation tool. We derived the guideline through a sequential process combining a literature review, user interviews, and a prioritization survey, and embedded it into a web-based tool for guided generation, quality checking, and iterative revision.

Our results indicate that users value explanations that are relevant, clear, concise, readable, and well structured, and that certain stylistic choices are perceived as inappropriate in this context. The prioritization study yielded 22 useful property manifestations, operational targets such as a preferred Flesch readability range of 40--49, and clear properties to avoid, including emojis and humorous or romantic writing style.

The tool evaluation suggests that guideline-based LLM support can improve perceived formulation quality and may offer efficiency benefits compared to manual formulation. Tool-supported writing was faster on average in our small-scale developer study and produced explanations that were rated significantly higher in overall satisfaction. Overall, the findings support treating explanation formulation as an engineering task with explicit quality criteria and using LLMs as guided authoring assistance under human control rather than as autonomous generators.

Future work should validate the approach across additional domains and explanation formats, and strengthen quality assurance for correctness and context fit in real-world workflows.

\section*{Acknowledgment}
This work was funded by the Deutsche Forschungsgemeinschaft (DFG, German Research Foundation) under Grant No. 470146331, project softXplainer (2026-2028).

\section*{Data Availability Statement}
\label{sec:datastatement}
All artifacts produced by or used in this study, including data, tool code, prompts, study materials, and analysis scripts, are publicly available on Zenodo~\cite{obaidi2026guidelineExpDataset}.

\bibliographystyle{IEEEtran}
\bibliography{references.bib}

\end{document}